# A new method for finding the minimum free energy pathway of ions and small molecule transportation through protein based on 3D-RISM theory and the string method

*Norio Yoshida**

Department of Chemistry, Graduate School of Science, Kyushu University

744, Motooka, Nishiku, Fukuoka, 819-0395 Japan

ABSTRACT

A new method for finding the minimum free energy pathway (MFEP) of ions and small molecule transportation through a protein based on the three-dimensional reference interaction site model (3D-RISM) theory combined with the string method has been proposed. The 3D-RISM theory produces the distribution function, or the potential of mean force (PMF), for transporting substances around the given protein structures. By applying the string method to the PMF surface, one can readily determine the MFEP on the PMF surface. The method has been applied to consider the $Na^+$ conduction pathway of channelrhodopsin as an example.

**Corresponding author**

* Norio Yoshida noriwo@chem.kyushu-univ.jp

# 1. Introduction

The transportation of ions and molecules through proteins is one of the most fundamental life phenomena to maintain living systems [1, 2]. An ion channel adjusts the ion concentration according to the difference of the chemical potential of both intra- and extracellular sides of the cell membrane. It is also related to the signal transduction processes and the conduction of electrical signals. While channels allow the transporting of ions or small molecules across the membrane only passively, a pump protein transports the substances actively across the membrane against chemical potential gradients. Because of active transport, pump proteins create membrane potentials or proton gradients between the intra- and extracellular sides of membranes.[3]

The characteristic of ion (or small molecule) transportation is expressed by the free energy profile or potential of mean force (PMF) through the protein. For example, the barrier on the PMF is related to the ion conduction flux. Therefore, the evaluation of the PMF is essential to understanding the mechanism of the transportation. However, computation of the PMF is a challenging issue for computational science because of the difficulties with evaluation of the PMF acting on the ions.

The PMF is defined as the potential that gives the average force taken for configurations of protein, water, and ions. Because ion transportation is highly coupled with the structural fluctuation of the protein and membrane, and the distribution of solvent species, the configurational integral for the PMF should include a sufficient number of configurations of both the protein structures and solvent distributions. Therefore, efficient sampling methods of the protein structures and the distributions of solvent, including water and ions, for the configuration integral are highly desirable. Many theoretical and computational approaches have been proposed to evaluate the PMF based on molecular dynamics (MD) simulations [4-6]. A nonphysical sampling such as umbrella sampling, metadynamics, and adaptive biasing force method is often employed to accelerate the MD sampling through the reaction coordinate of ion transportation [7-9]. However, these methods become inefficient when the number of reaction coordinates increases. Therefore, a method to determine the appropriate reaction coordinate is also important.

The difficulties in the sampling of the solvent distribution can be overcome by the three-dimensional reference interaction site model (3D-RISM) theory. The spatial distribution function (SDF) of the solvent by 3D-RISM is obtained through a complete ensemble average over the entire configuration space of solvent molecules, including ions in the thermodynamic limits [10-15]. Therefore, by applying 3D-RISM, one can determine the PMF surface of ions for the entire space prior to the reaction coordinate search for a given protein structure. To find the reaction coordinates on a given potential surface, the string method is effective [16-18] and has been applied successfully to various problems.

In this paper, we propose a new method for finding the minimum free energy pathway (MFEP) of ions (or small molecules) on the PMF surface obtained by 3D-RISM theory combined with the string method. This method consists of three steps. First, the SDF of ions is evaluated by 3D-RISM theory for given protein structures. The protein structures are assumed to be taken from the MD trajectory. (However, we employ the protein structure taken from the protein data bank (PDB) for simplicity in the present paper.) The PMF is readily obtained from the SDF. Second, the stationary position of the ions inside a channel is determined based on the SDF, where the SDF shows local maximum values. Third, the string method is used for the search of the MFEP that connects two stationary points on the PMF surface determined in the previous step [19-21]. A similar method based on the 3D-RISM theory and the elastic band method was proposed by Tanwar, Sindhikara, and co-workers, which is referred to as *RismPath* [22]. Because the method proposed in this paper is an extension of *RismPath,* we refer to the new method as *RismPath*/SM. The advantages of the string method over the elastic band method are higher stability and accuracy, which have been well discussed by E et al. [19]. The string method is a powerful tool to find an MFEP on a given surface that has been applied to various problems of finding the MFEP for a rare event involving a protein structure change [19, 23-25]. In contrast to those approaches, in the present method, the string method is used only for optimum route search that connects two points on the PMF surface with a given protein structure. In this study, the $Na^+$ conduction pathway of channelrhodopsin (ChR) is examined as an example to demonstrate the *RismPath*/SM method. In addition, to consider the effects of the protonation state of the amino acid residue on the pathway, both the protonated and unprotonated states of E136 and E140 have been examined. The MFEPs and the corresponding free energy profiles will be discussed.

## 2. Method

The computational procedure to determine the MFEP of the transporting substances through the protein is schematically depicted in Scheme 1. Although the present method can be applied to the MFEP search, not only for ions but also for small molecules, we focus on the ion case in this paper for simplicity.

First, to obtain the distribution function of the ions the 3D-RISM calculations should be conducted. Here, the target protein is regarded as a "solute" while the ions and water are treated as "solvents." The distribution function of solvent atom $\gamma$, $g_\gamma(\boldsymbol{r})$ is obtained by iteratively solving the 3D-RISM equation [10-12]

$$h_\gamma(\boldsymbol{r}) = \sum_{\alpha \in \text{solvent}} c_\alpha(\boldsymbol{r}) * X_{\alpha\gamma}(\boldsymbol{r} - \boldsymbol{r}'), \tag{1}$$

and the Kovalenko–Hirata closure equation [13, 26]

$$g_\gamma(\boldsymbol{r}) = \begin{cases} \exp(d_\gamma(\boldsymbol{r})) & d_\gamma(\boldsymbol{r}) < 0, \\ 1 + d_\gamma(\boldsymbol{r}) & d_\gamma(\boldsymbol{r}) \geq 0, \end{cases} \tag{2a}$$

$$d_\gamma(\boldsymbol{r}) = -\beta u_\gamma(\boldsymbol{r}) - h_\gamma(\boldsymbol{r}) - c_\gamma(\boldsymbol{r}), \tag{2b}$$

where $*$ denotes the convolution integral, $h_\gamma(\boldsymbol{r}) = g_\gamma(\boldsymbol{r}) - 1$ and $c_\gamma(\boldsymbol{r})$ are total and direct correlation functions, and $X_{\alpha\gamma}(\boldsymbol{r})$ is a bulk solvent susceptibility, which is obtained by solving the RISM for the bulk solvent system prior to the 3D-RISM calculation. $\beta = 1/k_\text{B}T$ is the thermodynamic beta. The solute–solvent interaction potential $u_\gamma(\boldsymbol{r})$ is given by

$$u_\gamma(\boldsymbol{r}) = \sum_{\alpha \in \text{solute}} \left\{ 4\varepsilon_{\alpha\gamma} \left[ \left(\frac{\sigma_{\alpha\gamma}}{r_{\alpha\gamma}}\right)^{12} - \left(\frac{\sigma_{\alpha\gamma}}{r_{\alpha\gamma}}\right)^{6} \right] + \frac{q_\alpha q_\gamma}{r_{\alpha\gamma}} \right\}, \tag{3}$$

where $r_{\alpha\gamma}$ is the distance between solute atom $\alpha$ and solvent atom $\gamma$. $\varepsilon_{\alpha\gamma}$, $\sigma_{\alpha\gamma}$, and $q_\alpha$ are the Lennard-Jones parameters and partial charge, respectively, with the usual meaning. Because the distribution functions obtained by 3D-RISM theory are obtained through a complete ensemble average over the entire configuration space of solvent molecules unlike MD, the distributions of ions inside a protein cavity or channel can be reproduced automatically. The PMF of solvent atom $\gamma$ is readily obtained by

$$\psi_\gamma(\boldsymbol{r}) = -\frac{1}{\beta} \log g_\gamma(\boldsymbol{r}). \tag{4}$$

It is noted that the PMF acting on the transporting species should include not only the solvent distribution also the distribution of the force from the fluctuating protein. However, in the present method, since we assumed the fixed protein structure, the PMF can be expressed as Eq. (4).

Because the distribution function, $g_\gamma(r)$, represents the probability of finding the solvent atom $\gamma$ at position $r$, the probable position to accommodate the ions can be specified based on the distribution functions. The *Placevent* algorithm is a simple algorithm explicitly to place solvent molecules using the 3D-RISM distribution function [27]. To place the solvent molecules, the population function is defined as

$$P_\gamma^{(0)}(r) = \rho_\gamma g_\gamma(r), \tag{5}$$

where $\rho_\gamma$ is a number density of solvent atom $\gamma$ in bulk solvent. The superscript of the population function denotes the number of iterations. The first explicit solvent atom is placed at the location with the highest population. After placing one explicit solvent atom, a new population function is obtained to maintain the total population of solvent atoms as

$$P_\gamma^{(i+1)}(r) = \begin{cases} P_\gamma^{(i)}(r) & r \notin v^{(i)} \\ 0 & r \in v^{(i)} \end{cases}, \tag{6}$$

where $v^{(i)}$ denotes the area inside the sphere of radius $\delta_i$ centered at $r_i$ where the $i$-th explicit solvent atom is placed. $\delta_i$ is determined to satisfy the condition:

$$\int_{r_i}^{r_i+\delta_i} P_\gamma^{(i)}(r)dr = 1, \tag{7}$$

once the explicit atom has been placed and the corresponding distribution has been reduced. The next explicit atom will be placed based on the new population function. These processes are repeated until a satisfactory number of explicit solvent atoms are placed.

The position of the explicit atom placed above may correspond to a stable or metastable position of the transporting substrate atoms. In the present study, the MFEP connecting two stationary points of ions is determined by the string method [19, 23, 24]. This is a distinguishing point of the new method proposed here from the *RismPath* method; therefore, we call the new method *RismPath*/String Method (*RismPath*/SM) hereafter. In the present study, we employed the zero-temperature string method for simplicity [23, 24]. It is noted that the finite temperature string method is also available, and it may be more effective in the case of a complicated PMF surface [20, 21]. In the string method, a string connecting initial and end points of the path is discretized using *N* discretization points, which are referred to as *images*. The initial guess of the images and the initial and end points of the path are taken from the results of the *Placevent* algorithm. Once the initial positions of images are given, the evolution of the string toward the MFEP is simulated by the optimization of each image on the PMF surface obtained by 3D-RISM calculation. The images on the string are evolved over time interval $\Delta t$ according to the PMF

$$\dot{\phi}_i = -\nabla \psi(\phi_i) \tag{8}$$

where $\dot{\phi}_i$ denotes the time derivative of coordinate of image $i$, $\phi_i$. Here, the PMF is a function of the coordinates of the ions, which can be regarded as collective variables for the string method. After the evolution of the images, the string is reparametrized, and one can obtain the new positions of the images. It is noted that the end points of strings are fixed during the MFEP search in this study. However, it is possible to use free moving end points, if required. These steps should be repeated until convergence to the MFEP.

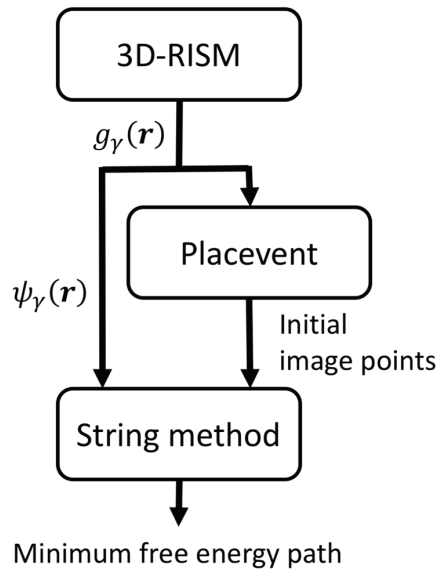

Scheme 1. Computational scheme of the *RismPath*/SM method to determine the MFEP on the given PMF surface of ions.

## 3. Computational details

To demonstrate the *RismPath*/SM method, we consider the Na$^+$ conduction pathway of ChR as an example. The protein structure is taken from the PDB entry code 3UG9, which is a crystal structure of the closed state of ChR (see Fig. 1A) [28]. Although the structure is a closed state, the pathway connecting the retinal-binding pocket to the extracellular region has been suggested, which is colored blue in Fig. 1B. In the present study, we used a fixed protein structure for simplicity. To consider the effects of the protonation state of the amino acid residue in the middle of the pathway, both the protonated and unprotonated states of E136 and E140 have been examined (see Table 1). We employed the CHARMM36 force field parameters for a standard amino acid [29] and the parameters for the Schiff base (SB) are taken from ref. [30]. The 3D-RISM calculations have been conducted using our own 3D-RISM program package [31]. NaCl solution (1 M) in ambient condition is assumed with the TIP3P water model at 298.15 K. The number of grid points in the 3D-RISM calculations was $256^3$ with a spacing of 0.5 Å. For the string method, we used a modified version of the string method program originally coded by Ren [19, 23]. Twenty-four images were prepared for each string, which corresponds to placing one image approximately every 1 Å. We employed Hermite interpolation to the PMF surface by 3D-RISM to obtain the PMF value at arbitrary coordinates for the string method.

Table 1. Model of protonation states of glutamate 136 (E136) and 140 (E140)

| Model | E136 | E140 |
|---|---|---|
| (a) | unprotonated | unprotonated |
| (b) | protonated | unprotonated |
| (c) | unprotonated | protonated |
| (d) | protonated | protonated |

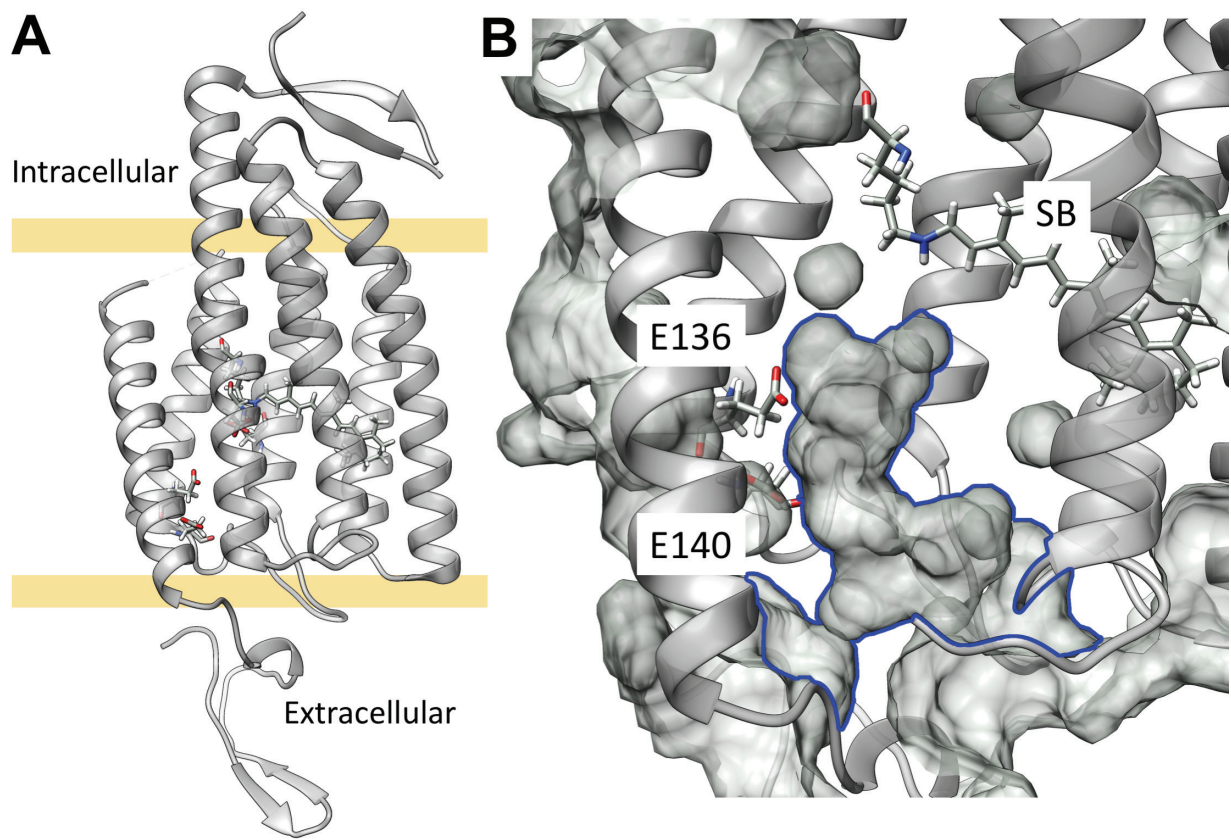

Figure 1. Structure of the closed state of C1C2 chimera ChR taken from PDB entry 3UG9. (A) is an overall structure and (B) depicts the candidate Na$^+$ conduction channel in blue.

## 4. Results and discussion

First, the 3D-RISM calculations for evaluating the distribution function of Na$^+$, Cl$^-$, and water have been conducted. The distribution functions of Na$^+$ and the oxygen of water around the ChR model (a) are depicted in Fig 2A and 2B, respectively. The isosurfaces of $g_{\mathrm{Na}^+}(\boldsymbol{r}) = 1.2$ and $g_\mathrm{O}(\boldsymbol{r}) = 1.2$ are colored green and red, respectively. The noteworthy distribution of Na$^+$ can be observed around E136 and E140, whereas the distribution of oxygen of water spreads throughout the channel. A small Na$^+$ distribution can also be found near SB.

Based on the distribution functions, the explicit water and Na$^+$ are placed by the *Placevent* algorithm. In Fig 2C, some of the explicit atoms are depicted, which will be employed as the start, intermediate, and end points of the string method. The Na$^+$ atom labeled "S" is coordinated to the side chain of D292, and that labeled "I" is coordinated to the side chain of E136. "$E_A$" and "$E_B$" are located near the exit on the extracellular side. The oxygen atoms of water labeled "$I_A$" and "$I_B$" are located in I and $E_A$, I and $E_B$, respectively.

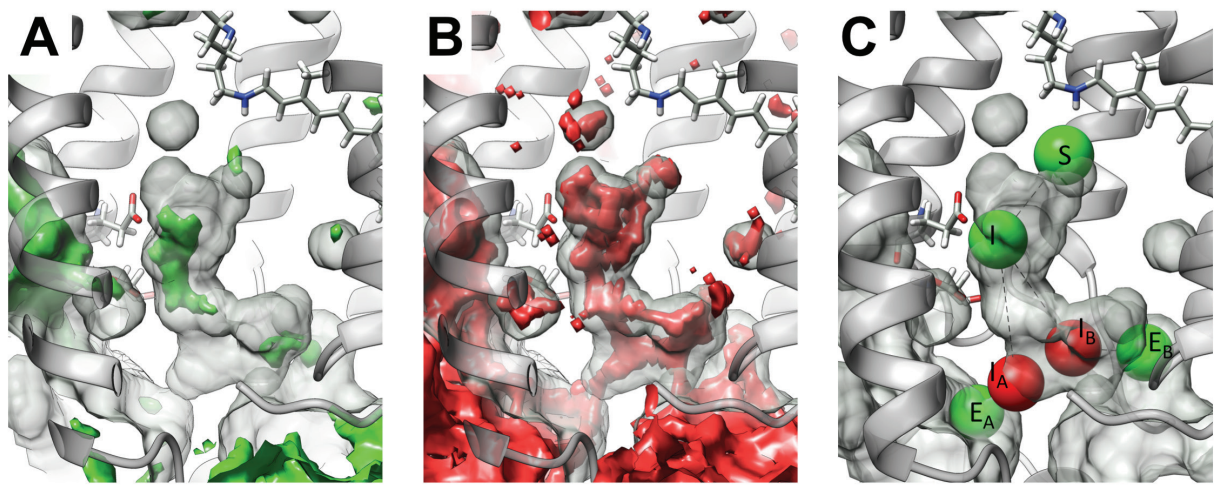

Figure 2. The distribution function of (A) Na$^+$ and (B) oxygen of water of model (a). The isosurfaces of distribution functions $g(r) \geq 1.2$ are plotted in green and red for Na$^+$ and oxygen, respectively. (C) The explicit Na$^+$, green, and water oxygen, red, placed using *Placevent*.

Here, we assumed two routes of the Na$^+$ conduction pathway, namely, S–$E_A$ and S–$E_B$, which are referred to as routes A and B, respectively. To prepare the initial guess of the images on the string connecting S and $E_A$ or $E_B$, the atoms "I," "$I_A$," and "$I_B$" are used as the intermediate sites. The images between S–I, I–$I_A$, $I_A$–$E_A$, I–$I_B$, and $I_B$–$E_B$ are generated equidistantly in a straight line connecting the two sites. Five individual computations of the string method for the string S–I, I–$I_A$, $I_A$–$E_A$, I–$I_B$, and $I_B$–$E_B$ were conducted. By combining these short strings, the strings connecting S–$E_A$ and S–$E_B$ are constructed. In other words, the guess for images connecting S–$E_A$ (or S–$E_B$) is taken from the results of the string method of S–I plus I–$I_A$ plus $I_A$–$E_A$ (or S–I plus I–$I_B$ plus $I_B$–$E_B$).

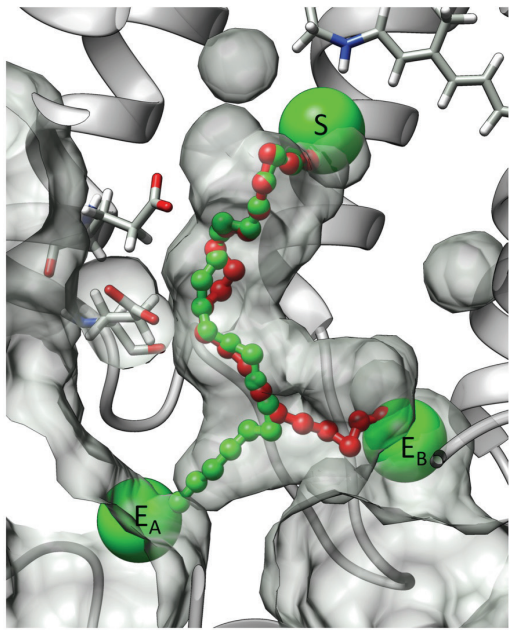

Figure 3. The MFEPs connecting site S to $E_A$ and $E_B$. The small spheres in green and red denote the images on the MFEPs of routes A and B, respectively.

In Fig. 3, the results of the string method for the ChR model (a) for both routes A and B are shown. The small spheres colored green and red denote the images on the MFEPs of routes A and B, respectively. Each image is placed at 1.09 Å and 1.11 Å for routes A and B, respectively. The PMF profiles along the MFEPs are plotted against the images in Fig. 4, where the image 0 corresponds to the atom "S" while the image 25 is "$E_A$" (or "$E_B$"). At a glance, a peak can be found at images 2 to 3. This is because the positively charged K132 side chain is close to the images. The images 6 to 13 are stabilized by the negatively charged E136 and E140. The high barrier that appeared at around image 23 is attributed to steric hindrance by N82 and Coulomb repulsion by K154. In the case of route B, the conspicuous barrier is observed around images 23 and 24, which is produced by the steric hindrance by K154 and P273. Therefore, route B may be an unfavorable path for Na$^+$ conduction of this ChR structure.

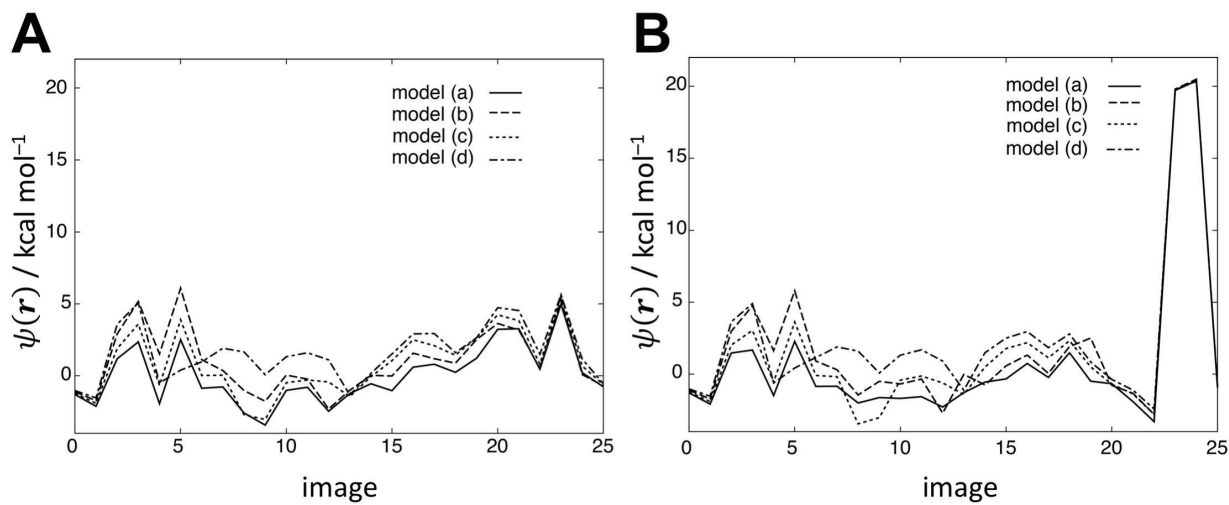

Figure 4. The PMF profiles along the MFEPs of (A) route A and (B) route B. The PMF values at each image are plotted. The images 0 and 25 correspond to sites S and $E_A$ (or $E_B$), respectively.

To consider the effects of the protonation state of E136 and E140, models that have different protonation states of E136 and E140 have been examined. The MFEPs for models (b), (c), and (d) were determined by the string method, where the MFEP for model (a) was employed as an initial guess. The PMF profiles along the MFEPs are plotted against the images in Fig. 4. By the protonation of glutamic acid, the attractive Coulomb interaction between the carboxyl group and Na$^+$ is reduced and the barriers of PMF appear around images 6 and 16. In particular, the doubly protonated model, model (d), shows drastic changes in the PMF profiles for both routes A and B. In Fig. 5, the MFEPs of the models (a) and (b) for route A are compared. As can be seen in the figure, the MFEP of model (b) slightly changed around E136 and E140 to circumvent the residues while almost all other parts are conserved.

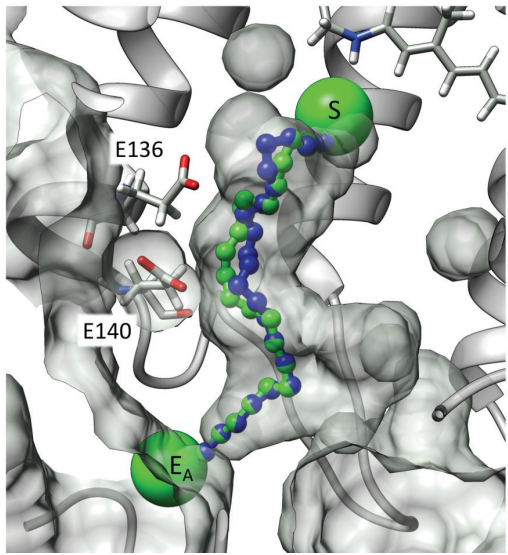

Figure 5. The MFEPs of route A for model (a), green, and model (d), blue.

To assess the numerical stability of the method, we performed additional calculations with different parameters for the grid spacing of 3D-RISM and the number of images for the string method. Here, we employed a fine grid spacing, $\Delta r = 0.25$ Å, and a large image number, $N_i = 49$, for comparison. In Fig. 6, the MFEPs of route A with the different parameters are depicted. At a glance, all the MFEPs show good agreement with each other. Relatively large differences can be seen around E136. The distance between the corresponding images is 0.5 Å at largest. In Fig. 7, the PMF values along the MFEPs with the different parameters are compared. In Fig. 7A, the grid spacing dependence of PMF is considered. Although a relatively large discrepancy, ~3 kcal mol$^{-1}$, is observed at 24, the PMF values are in good agreement with the fine grid case. Therefore, the grid spacing $\Delta r = 0.50$ Å often used in the 3D-RISM calculation may be acceptable for the MFEP search. When $\Delta r = 0.25$ Å is employed, we should use $512^3$ grid points to maintain the solvent box size. Even considering that the computational cost of the 3D-RISM calculation is almost proportional to the number of grid points, for the accurate evaluation of the barrier height, the adoption of the fine grid spacing should be considered. In Fig. 7B, the PMFs with different numbers of image points are compared. Although the PMF profiles show good agreement with each other, some of the barriers are slightly underestimated in the $N_i = 24$ case. This is because the barrier is located between two images in this case. The computational cost for $N_i = 49$ is larger than for $N_i = 24$. However, because the computational cost for the string method in the *RismPath*/SM is minor, it is recommended to employ a large number of image points.

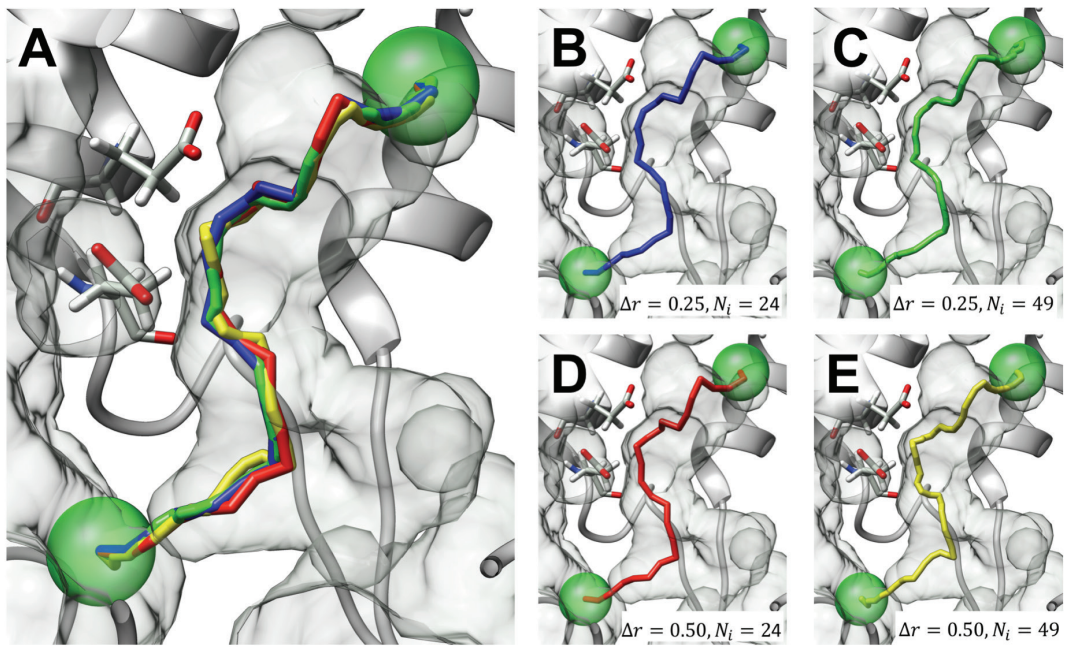

Figure 6. The MFEPs of route A with different parameters. Panel (A) depicts all the MFEPs with the different parameters whereas panels (B) to (E) depict the MFEPs individually. The parameters of the grid spacing, $\Delta r$, and the number of images, $N_i$, employed here are indicated in the bottom of each panel.

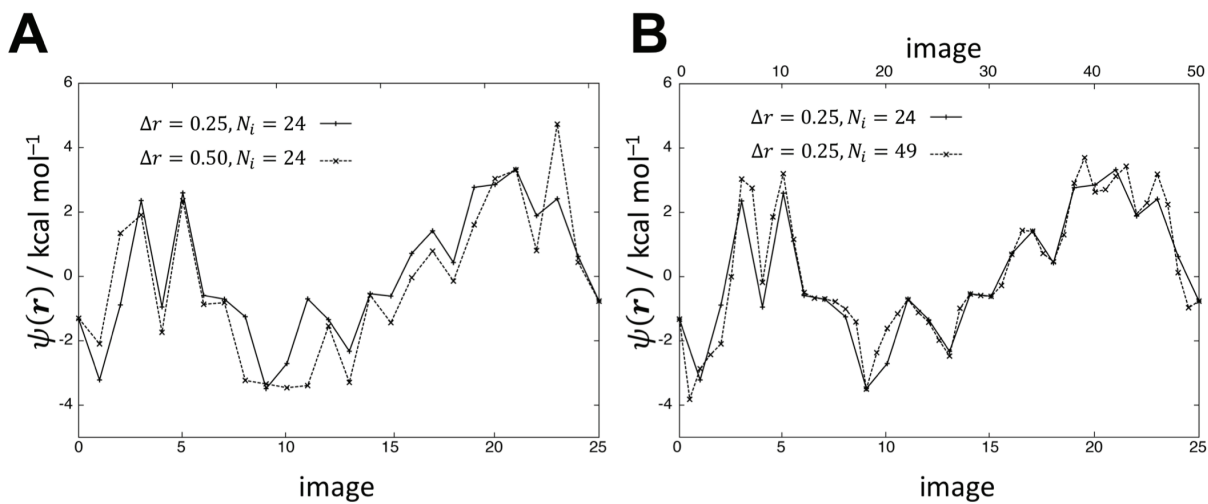

Figure 7. Comparison of the PMF along the MFEPs with different parameters. The PMF values obtained by the different grid spacing are compared in panel (A), and the PMF values with different numbers of images are compared in panel (B). In panel (B), lower and upper horizontal axes correspond to $N_i = 24$ and $N_i = 49$, respectively. The images 0 and 25 (or 50) correspond to sites S and $E_A$, respectively.

## 5. Summary

In this letter, we have proposed a new method for finding the MFEP of ions and small molecule transportation through a channel protein based on the 3D-RISM theory combined with the string method, which is referred to as the *RismPath*/SM. Thanks to the 3D-RISM theory for evaluating the PMF of ions, the difficulties with sampling the ion and solvent distributions can be overcome. In the present study, we employed the zero-temperature string method for simplicity. It is noted that the finite temperature string method is also available, and it may be more effective in the case of a complicated PMF surface.

We applied the *RismPath*/SM to evaluate the MFEP of the Na⁺ conduction pathway of the ChR closed state as an example. The method presented here is focused on the improvement of solvent distribution sampling for a given fixed protein structure. As expected, the structural fluctuation of the channel protein plays an important role in molecular transportation. Therefore, a combination with a method that can take the structural fluctuation of a protein into account, such as MD simulation, is strongly desired for highly accurate analysis. The self-consistent type of hybrid method of the 3D-RISM theory and MD simulation is suitable for this purpose [32, 33]. In the method, the protein (and membrane, if any) structure is treated by MD simulation whereas the distributions of the ions and solvent molecules are evaluated using 3D-RISM theory. The combination of the MD/3D-RISM method with the efficient sampling methods of the protein structure is useful for providing the PMF surface to the *RismPath*/SM. Such studies with the *RismPath*/SM are in progress.

**Acknowledgments**

This work was supported by Grants-in-Aid [16H00842, 16K05519] from MEXT, Japan. The author is grateful to Dr. Daniel J. Sindhikara (Schrodinger Inc.), Prof. Shigehiko Hayashi (Kyoto University), and Prof. Haruyuki Nakano, (Kyushu University) for invaluable discussions. The original program of the string method was taken from Dr. Weiqing Ren's website (http://www.math.nus.edu.sg/~matrw/string/index.html). Molecular graphics and analyses were performed with the UCSF Chimera package [34].

**References**

[1] B. Hill,ed., Ion channels of excitable membranes. 3rd edn. Sinauer Associates Inc., Sunderland, MA, 2001.
[2] J.N.C. Kew, C.H. Davies ed., Ion channels from structure to function. Oxford university press, Oxford, 2010.
[3] H.R. Zhekova, V. Ngo, M.C. da Silva, D. Salahub, S. Noskov, Coord. Chem. Rev. 345 (2017) 108.

[4] C. Bossa, M. Anselmi, D. Roccatano, A. Amadei, B. Vallone, M. Brunori, A. Di Nola, Biophys. J. 86 (2004) 3855.
[5] G. Bussi, A. Laio, M. Parrinello, Phys. Rev. Lett. 96 (2006) 090601.
[6] Y. Nishihara, S. Hayashi, S. Kato, Chem. Phys. Lett. 464 (2008) 220.
[7] J. Kottalam, D.A. Case, J. Am. Chem. Soc. 110 (1988) 7690.
[8] S. Kumar, J.M. Rosenberg, D. Bouzida, R.H. Swendsen, P.A. Kollman, J. Comput. Chem. 13 (1992) 1011.
[9] C. Chipot, J. Henin, J. Chem. Phys. 123 (2005).
[10] D. Beglov, B. Roux, J. Chem. Phys. 104 (1996) 8678.
[11] D. Beglov, B. Roux, J. Phys. Chem. B 101 (1997) 7821.
[12] A. Kovalenko, F. Hirata, Chem. Phys. Lett. 290 (1998) 237.
[13] A. Kovalenko, F. Hirata, J. Chem. Phys. 110 (1999) 10095.
[14] N. Yoshida, J Chem Inf Model 57 (2017) 2646.
[15] S. Phongphanphanee, N. Yoshida, F. Hirata, Curr. Pharm. Des. 17 (2011) 1740.
[16] A.C. Pan, D. Sezer, B. Roux, J Phys Chem B 112 (2008) 3432.
[17] L. Maragliano, A. Fischer, E. Vanden-Eijnden, G. Ciccotti, J. Chem. Phys. 125 (2006) 024106.
[18] C. Zhao, S.Y. Noskov, PLoS Comput Biol 9 (2013) e1003296.
[19] W.N. E, W.Q. Ren, E. Vanden-Eijnden, J. Chem. Phys. 126 (2007) 164103.
[20] W. Ren, E. Vanden-Eijnden, P. Maragakis, W. E, J. Chem. Phys. 123 (2005) 134109.
[21] W. E, W. Ren, E. Vanden-Eijnden, J. Phys. Chem. B 109 (2005) 6688.
[22] A.S. Tanwar, D.J. Sindhikara, F. Hirata, R. Anand, ACS Chem Biol 10 (2015) 698.
[23] E. Weinan, W.Q. Ren, E. Vanden-Eijnden, Phys Rev B 66 (2002) 052301.
[24] L. Maragliano, A. Fischer, E. Vanden-Eijnden, G. Ciccotti, J. Chem. Phys. 125 (2006) 024106.
[25] L. Maragliano, G. Cottone, G. Ciccotti, E. Vanden-Eijnden, J. Am. Chem. Soc. 132 (2010) 1010.
[26] A. Kovalenko, F. Hirata, J. Phys. Chem. B 103 (1999) 7942.
[27] D.J. Sindhikara, N. Yoshida, F. Hirata, J Comput Chem 33 (2012) 1536.
[28] H.E. Kato, F. Zhang, O. Yizhar, C. Ramakrishnan, T. Nishizawa, K. Hirata, J. Ito, Y. Aita, T. Tsukazaki, S. Hayashi, P. Hegemann, A.D. Maturana, R. Ishitani, K. Deisseroth, O. Nureki, Nature 482 (2012) 369.
[29] R.B. Best, X. Zhu, J. Shim, P.E.M. Lopes, J. Mittal, M. Feig, A.D. MacKerell, J. Chem. Theory Comput. 8 (2012) 3257.
[30] S. Hayashi, I. Ohmine, J. Phys. Chem. B 104 (2000) 10678.
[31] Y. Maruyama, N. Yoshida, H. Tadano, D. Takahashi, M. Sato, F. Hirata, J. Comput. Chem. 35 (2014) 1347.
[32] T. Luchko, S. Gusarov, D.R. Roe, C. Simmerling, D.A. Case, J. Tuszynski, A. Kovalenko, J. Chem. Theory Comput. 6 (2010) 607.
[33] T. Miyata, F. Hirata, J. Comput. Chem. 29 (2008) 871.
[34] E.F. Pettersen, T.D. Goddard, C.C. Huang, G.S. Couch, D.M. Greenblatt, E.C. Meng, T.E. Ferrin, J. Comput. Chem. 25 (2004) 1605.